\documentclass[prd,preprint,showpacs,groupedaddress]{revtex4-1}
\usepackage{amsmath}
\usepackage{amsfonts}
\usepackage{amssymb}
\usepackage{geometry}
\usepackage{graphicx}
\usepackage{natbib}
\usepackage[english]{babel}
\usepackage{graphicx}
\usepackage{epstopdf}
\usepackage{subfigure}
\usepackage{caption}
\usepackage{multirow}
\usepackage{indentfirst}
\usepackage{diagbox}
\usepackage{mathrsfs}
\usepackage{booktabs}
\usepackage[colorlinks,linkcolor=blue,anchorcolor=,citecolor=blue]{hyperref}
\usepackage{amsthm}
\usepackage{float}
\numberwithin{equation}{section}
\numberwithin{figure}{section}

\begin{document}
    \title{Euler-Heisenberg black hole surrounded by perfect fluid dark matter}
    \author{Shi-Jie Ma}
    \affiliation{Lanzhou Center for Theoretical Physics, Key Laboratory of Theoretical Physics of Gansu Province, Lanzhou University, Lanzhou, Gansu 730000, China}
    
    \author{Rui-Bo Wang}
	\affiliation{Lanzhou Center for Theoretical Physics, Key Laboratory of Theoretical Physics of Gansu Province, Lanzhou University, Lanzhou, Gansu 730000, China}
	
	\author{Jian-Bo Deng}
	\email[Email: ]{dengjb@lzu.edu.cn}
	\affiliation{Lanzhou Center for Theoretical Physics, Key Laboratory of Theoretical Physics of Gansu Province, Lanzhou University, Lanzhou, Gansu 730000, China}
	
	\author{Xian-Ru Hu}
	\email[Email: ]{huxianru@lzu.edu.cn}
	\affiliation{Lanzhou Center for Theoretical Physics, Key Laboratory of Theoretical Physics of Gansu Province, Lanzhou University, Lanzhou, Gansu 730000, China}
	
	\date{\today}
    \begin{abstract}
        A generation method of new metric in the case of static spherically symmetric space-time is derived. Using this approach, we construct a metric which describes Euler-Heisenberg black hole surrounded by perfect fluid dark matter and investigate its optical and thermodynamic properties. We found that radius of shadow will increase with the increase of dark matter effect, and more strong dark matter will diminish the light intensity of accretion disk generally. Moreover, in thermodynamics, when quantum electrodynamic parameter is positive, there will be a critical value of dark matter parameter, which determine the number of black hole's critical points.
    \end{abstract}
   
    \maketitle
    \section{Introduction}\label{sec1}
        Euler-Heisenberg (EH) theory is a kind of nonlinear electrodynamics, which takes quantum electrodynamics (QED) into account~\cite{EHtheory}. Based on EH theory, scientists obtained charged black holes called 'EH black holes'~\cite{EHsolution}, whose optics and thermodynamics have been widely studied. EH black hole's shadow and luminosity of accretion flow aren't sensitive to QED effect, which possibly indicates one-loop correction to QED hardly works in optics~\cite{QED_EH}. In contrast, QED correction will bring a transparent impact in thermodynamics compared to RN black hole. For instance, QED parameter will crucially determine the number of critical points~\cite{EHbh2,EHbh3} and it will cause a small deviation on the critical ratio~\cite{EHbH1}. In addition, high-order QED correction has been considered as well~\cite{EHbh4}. On the other hand, it is believed that dark matter could potentially explain the expanding universe. As a special model of dark matter, perfect fluid dark matter (PFDM) is an ideal model that satisfies equation of state $p/\rho=\epsilon$, leading to a static spherically space-time described by the power function or the logarithmic function~\cite{DMI,DMII,DMIII}. The possibility of PFDM in a phantom field constrained by motions of stars in spiral galaxy was studied in~\cite{DMIV,DMV}. Especially, PFDM inducing logarithmic form of space-time metric has attracted more interest of physicists~\cite{DMV,DMII,DMVI}. Studying the properties of black holes in the presence of PFDM will uncover the influence of dark matter on black hole physics and help to understand their mechanism in theory of gravity~\cite{DMI,DMVI,DMII}.
        
        Optics and thermodynamics are main subjects in black hole physics and already have very mature research process. Due to black hole's huge gravitational pull, the  light surrounding it will be significantly bent, leading to fascinating optical phenomena, with the shadow and image of accretion disk being most captivating~\cite{image1,image2,image3,shadow,isco}. In thermodynamics, since Hawking and Page firstly studied the thermodynamics of black hole in anti-de Sitter (AdS) space-time~\cite{Hawking1}, there are many significant researches published~\cite{RN-Ads1,RN-Ads2,Kastor,Dolan}. Generally speaking, the variation of cosmological constant in extended phase space will lead to more interesting results.
        
        This paper aims to take EH black hole into PFDM background and study its optical nature and thermodynamics. To obtain the solution of field equation more conveniently, this article will also provide a method to generate the new metric in static spherically symmetric space-time. Looking forward, if there are related experiments in future, our research will help to distinguish whether QED effect and dark matter work. The article is organized as follows. In Sect.~\ref{sec2}, we discuss the generation of new metric and give its corresponding conditions mathematically. In Sect.~\ref{sec3}, the metric of EH black hole surrounded by PFDM is obtained. In Sect.~\ref{sec4}, we calculate the photon orbit and plot the image of thin accretion disk using Novikov-Thorne model. The results with different parameters and effect of Doppler shift will be shown. In Sect.~\ref{sec5}, thermodynamics of EH-AdS black hole surrounded by PFDM will be investigated, especially its criticality. The conclusion and outlook will be given in Sect.~\ref{sec6}. We always set $G=c=k=\hbar=1$.

    \section{Generation of new metric}\label{sec2}
        Consider a action
        \begin{equation}\label{eq2_1}
    		S=\int \sqrt{-g}\left({\frac{R}{16 \pi}+\sum_{i}\mathcal{L}^{\left(i\right)}}\right)d^{4}x,
    	\end{equation}
         where $g$ is determinant of the metric $g_{\mu\nu}$, $R$ is scalar curvature and $\mathcal{L}^{\left(i\right)}$ is Lagrangian, which represents $i^{\rm{th}}$ effect, like electromagnetic field, dark matter, etc. We always suppose every $\mathcal{L}^{\left(i\right)}$ has its corresponding static spherically symmetric solution and we only discusss static spherically symmetric space-time in this article. The coordinate system takes spherical coordinate $\left(t,r,\theta,\phi\right)$, where $t\in\mathbb{R}$, $r>0$, $\theta\in\left[0,\pi\right]$ and $\phi \in \mathbb{R}$.
        
        By varying the Eq.~\eqref{eq2_1} with the respect to the $g^{\mu\nu}$, one could get the gravitational field equation,
        \begin{equation}\label{eq2_2}
            R_{\mu\nu}-\frac{1}{2}g_{\mu\nu}R=8\pi T_{\mu\nu},
        \end{equation}
        where $R_{\mu\nu}$ is Ricci tensor and $T_{\mu\nu}$ is energy momentum tensor, which is variation of $\mathcal{L}^{\left(i\right)}$:
         \begin{equation}\label{eq2_3}
        \begin{aligned}
            T_{\mu\nu}&=-\frac{2}{\sqrt{-g}}\frac{\delta\left(\sqrt{-g}\sum_{i}\mathcal{L}^{\left(i\right)}\right)}{\delta g^{\mu\nu}}\\
            &=\sum_{i} -\frac{2}{\sqrt{-g}}\frac{\delta\left(\sqrt{-g}\mathcal{L}^{\left(i\right)}\right)}{\delta g^{\mu\nu}}\\
            &=\sum_{i}T_{\mu\nu}^{\left(i\right)},
            \end{aligned}
        \end{equation}
        
       Now consider a specific circumstance that Lagrangians $\mathcal{L}^{\left(i\right)}$ in Eq.~\eqref{eq2_1} are independent. Generally, the form of $\mathcal{L}^{\left(i\right)}$ is
        \begin{equation}\label{eq2_4}
            \mathcal{L}^{\left(i\right)}=\mathcal{L}^{\left(i\right)}\left(g^{\mu\nu}, \partial_\gamma g^{\mu\nu},A_{\nu}^{\left(i\right)},\partial_\mu A_{\nu}^{\left(i\right)}\right),
        \end{equation}
        where $A_{\mu}^{\left(i\right)}$ describes the $i^{\rm{th}}$ effect. We just consider the first derivative of $A_{\mu}^{\left(i\right)}$. As an example, for electromagnetic field, it is four-potential. The independence of $\mathcal{L}^{\left(i\right)}$ means that $\mathcal{L}^{\left(i\right)}$ has no relationship with $A_{\mu}^{\left(j\right)}$ if $i\neq j$:
        \begin{equation}\label{eq2_5}
        	\frac{\delta \mathcal{L}^{\left(i\right)}}{\delta A_{\mu}^{\left(j\right)}}=0,~i\neq j.
        \end{equation}
        Theoretically, if two effects are coupling, there should be a Lagrangian describing their interaction:
        \begin{equation}\label{eq2_6}
            \mathcal{L}^{\left(i,j\right)}=\mathcal{L}^{\left(i,j\right)}\left(g^{\mu\nu},\partial_\gamma g^{\mu\nu},A_{\nu}^{\left(i\right)},A_{\nu}^{\left(j\right)},\partial_\mu A_{\nu}^{\left(i\right)},\partial_\mu A_{\nu}^{\left(j\right)}\right).
        \end{equation}
        Only combining $\mathcal{L}^{\left(i,j\right)}$, $\mathcal{L}^{\left(i\right)}$ and $\mathcal{L}^{\left(j\right)}$ as one term could deal with this problem.

        For our later discussion, we use $g^{\mu\nu}$ to raise index of Eq.~\eqref{eq2_2}:
        \begin{equation}\label{eq2_7}
            R_{\mu}^{\nu}-\frac{1}{2}\delta_{\mu}^{\nu}R=8\pi \sum_{i}T_{\mu}^{\nu\left(i\right)},
        \end{equation}
        where $T_{\mu}^{\nu\left(i\right)}=g^{\nu\alpha}T_{\mu\alpha}^{\left(i\right)}$. Assume $T_{0}^{{0}\left(i\right)}$ is independent from space-time itself. It means after solving the field equation of $A_{\mu}^{\left(i\right)}$:
        \begin{equation}\label{eq2_8}
            \frac{\delta}{\delta A_{\mu}^{\left(i\right)}}\int \sqrt{-g}\mathcal{L}^{\left(i\right)}d^4x=0,
        \end{equation}
        substituting the solution $A_{\mu}^{0\left(i\right)}$ and metric only $\mathcal{L}^{\left(i\right)}$ exists,
        \begin{equation}\label{eq2_9}
            g_{\mu\nu}^{\left(i\right)}=diag\left(-\left(1+f_{i}\left(r\right)\right),\frac{1}{1+f_{i}\left(r\right)},r^2,r^2\sin{\theta}\right)
        \end{equation}
        into $T_{0}^{0\left(i\right)}$, the result should be independent from the $f_{i}\left(r\right)$, which means
        \begin{equation}\label{eq2_10}
             \frac{\delta}{\delta f_{i}\left(r\right)}\left(T_{0}^{0\left(i\right)}\bigg|_{A_{\mu}^{0\left(i\right)},g_{\mu\nu}^{\left(i\right)}}\right)=0.
        \end{equation}
      
        Now the solution of Eq.~\eqref{eq2_7} could be discussed under above two requirements.
       
        One could get $A_{\mu}^{\left(i\right)}$ from
        \begin{equation}\label{eq2_12}
            \frac{\delta}{\delta A_{\mu}^{\left(i\right)}}\int \sqrt{-g}\sum_{k}\mathcal{L}^{\left(k\right)}d^4x=0.
        \end{equation}
        According to independence of $\mathcal{L}^{\left(i\right)}$, this equation is equivalent to Eq.~\eqref{eq2_8}. Assume $A_{\mu}^{0\left(i\right)}$ is the solution of Eq.~\eqref{eq2_8} and the solution of Eq.~\eqref{eq2_7} is
        \begin{equation}\label{eq2_13}
            g_{\mu\nu}^{\left(total\right)}=diag\left(-\left(1+f\left(r\right)\right),\frac{1}{1+f\left(r\right)},r^2,r^2\sin{\theta}\right).
        \end{equation}
        Substitute $g_{\mu\nu}^{\left(total\right)}$ into Eq.~\eqref{eq2_7}, nonzero term is
        \begin{equation}\label{eq2_14}
    		\frac{f\left(r\right)+rf'\left(r\right)}{r^2}=8\pi\sum_{i}T_{0}^{0\left(i\right)}\Bigg|_{A_{\mu}^{0\left(i\right)},g_{\mu\nu}^{\left(total\right)}}.
        \end{equation}
        In the same way,
        \begin{equation}\label{eq2_15}
            \frac{f_{i}\left(r\right)+rf_{i}'\left(r\right)}{r^2}=8\pi T_{0}^{0\left(i\right)}\Bigg|_{A_{\mu}^{0\left(i\right)},g_{\mu\nu}^{\left(i\right)}}.
        \end{equation}
        Use Eq.~\eqref{eq2_10}, we can surely replace $g_{\mu\nu}^{\left(i\right)}$ in Eq.~\eqref{eq2_15} with $g_{\mu\nu}^{\left(total\right)}$. Furthermore, according to Eq.~\eqref{eq2_14} and Eq.~\eqref{eq2_15}, one get
        \begin{equation}\label{eq2_17}
            f\left(r\right)=\frac{C}{r}+\sum_{i}f_{i}\left(r\right),
        \end{equation}
        where $C$ is integration constant to be determined.
        
        The solution of Eq.~\eqref{eq2_7} is finally derived:
        \begin{equation}\label{eq2_18}
            \begin{aligned}
            ds^2=-\left(1+f\left(r\right)\right)dt^2+&\frac{1}{1+f\left(r\right)}dr^2+r^2d\Omega^2,\\
            f\left(r\right)=\frac{C}{r}+&\sum_{i}f_{i}\left(r\right),
            \end{aligned}
        \end{equation}
        where $d\Omega^2=d\theta^2+\sin^2{\theta}d\phi^2$ is Euclidean line element on two-dimensional sphere.
        
        Here, it's necessary to point out that above discussion is just a mechanical deduction. In practice, if $T_{0}^{0\left(i\right)}$ in case of various interactions can be replaced with $T_{0}^{0\left(i\right)}$ when there is only one interaction, which means function of space-time itself has no influence on form of $T_{0}^{0\left(i\right)}$, then one can draw a conclusion that the metric will simply meet with Eq.~\eqref{eq2_17}. Just in mathematical way, Eq.~\eqref{eq2_5} and Eq.~\eqref{eq2_10} can insure it.
       
        It's reasonable to propose $f_{i}\left(r\right)$ in Eq.~\eqref{eq2_9} as the influence of $\mathcal{L}^{\left(i\right)}$ on the flat space-time. So, if many interactions have no coupling with each other, this impact on the space-time can get simply superimposed. Moreover, Schwawrzschild-like term $C/r$ in Eq.~\eqref{eq2_17} will not hinder the superimposition because $C/r$ has no contribution to energy momentum tensor.
    \section{EH Black Hole surrounded by PFDM}\label{sec3}
        In this section, we use the method in the previous section to construct a space-time, which describes a EH black hole surrounded by PFDM. The optical properties and thermodynamics will be investigated in Sect.~\ref{sec4} and Sect.~\ref{sec5}.
        
        Firstly,~\cite{EHsolution,QED_EH} provides the line element of EH black hole,
        \begin{equation}\label{eq3_1}
            \begin{aligned}
            ds^2=-\left(1+f_{1}\left(r\right)\right)dt^2&+\frac{1}{1+f_{1}\left(r\right)}dr^2+r^2d\Omega^2,\\
            f_{1}\left(r\right)=-\frac{2M}{r}&+\frac{Q^2}{r^2}-\frac{aQ^4}{20r^6},
            \end{aligned}
        \end{equation}
        its corresponding energy momentum tensor can be found from~\cite{EHbH1},
        \begin{equation}\label{eq3_2}
            T_{t}^{t}=\frac{1}{4\pi}\left(-\frac{Q^2}{2r^4}+\frac{aQ^4}{8r^8}\right),
        \end{equation}
        where $M$ is the mass of the black hole, $Q$ is the electric charge of the black hole and $a$ is the  QED parameter which is used  to measure the strength of the QED correction.
        
        Secondly, we reference the energy momentum tensor of PFDM mentioned in~\cite{DMVI},
        \begin{equation}\label{eq3_3}
           T_{t}^{t}=\frac{\alpha}{8\pi r^3},
        \end{equation}
        where $\alpha$ is the dark matter parameter. Via substituting Eq.~\eqref{eq3_3} into Eq.~\eqref{eq2_14}, one get the line element of space-time only PFDM exists,
        \begin{equation}\label{eq3_4}
            \begin{aligned}
            ds^2=-\left(1+f_{2}\left(r\right)\right)&dt^2+\frac{1}{1+f_{2}\left(r\right)}dr^2+r^2d\Omega^2,\\
            f_{2}\left(r\right)&=\frac{\alpha}{r}\ln{\frac{r}{\left|\alpha\right|}}.
            \end{aligned}
        \end{equation}
        
        Considering that dark matter won't participate with the electromagnetic interaction and it is obvious that above two energy momentum tensors are both independent from metric function $f\left(r\right)$ to be determined. According to generation method studied in Sect.~\ref{sec2}, the line element of EH black hole surrounded by PFDM is derived
        \begin{equation}\label{eq3_5}
            \begin{aligned}
            ds^2=&-g\left(r\right)dt^2+\frac{1}{g\left(r\right)}dr^2+r^2d\Omega^2,\\
            g\left(r\right)=1&-\frac{2M}{r}+\frac{Q^2}{r^2}-\frac{aQ^4}{20r^6}+\frac{\alpha}{r}\ln\left(\frac{r}{\left|\alpha\right|}\right).
            \end{aligned}
        \end{equation}
        Please notice that integration constant $C$ has been chosen as zero to insure Eq.~\eqref{eq3_5} can degenerate to EH black hole by choosing $\alpha$ as zero.
       	
       	For convenience, we uniformly use $M=1$ in the figures in Sect.~\ref{sec4}.
    \section{Optical properties}\label{sec4}
    	In this section, we will study the optical properties of EH black hole surrounded by PFDM, including the photon orbit and image of thin accretion disk.
    \subsection{Geodesic equation and photon orbit}\label{sec4_1}
    	Consider a particle moving in this space-time
    	\begin{equation}\label{eq4_1}
    		ds^2=-g\left(r\right)dt^2+\frac{dr^2}{g\left(r\right)}+r^2d\Omega^2,
    	\end{equation}
         
         A particle's Lagrangian is
    	\begin{equation}\label{eq4_2}
    		\mathcal{L}=\frac{1}{2}g_{\mu\nu}\dot{x}^{\mu}\dot{x}^{\nu},
    	\end{equation}
    	where $\dot{x}^{\mu}=dx^{\mu}/d\lambda$ is four-velocity of particle, $\lambda$ is proper time $\tau$ for physical particle and affine parameter for photon.
    	For simplicity, we choose $\theta=\pi/2$, $\dot{\theta}=0$, which means particle only moves in equatorial plane.
    	Euler-Lagrange equation gives two conserved quantities,
    	\begin{equation}\label{eq4_3}
    		E=-\frac{\partial \mathcal{L}}{\partial \dot{t}}=-g_{tt}\dot{t},
    	\end{equation}
    	\begin{equation}\label{eq4_4}
    		L=\frac{\partial \mathcal{L}}{\partial \dot{\phi}}=g_{\phi\phi}\dot{\phi},
    	\end{equation}
    	where $E$ and $L$ represent the energy and angular momentum of particle.
    	According to Eq.~\eqref{eq4_2}, Eq.~\eqref{eq4_3} and Eq.~\eqref{eq4_4},
    	\begin{equation}\label{eq4_7}
    		\dot{r}^{2}=E^{2}-g\left(r\right)\frac{L^{2}}{r^{2}}+2g\left(r\right)\mathcal{L}.
    	\end{equation}    	
    	If use Eq.~\eqref{eq4_4} and Eq.~\eqref{eq4_7} to eliminate parameter $\lambda$,
    	\begin{equation}\label{eq4_8}
    		\left(\frac{dr}{d\phi}\right)^{2}=r^4\left(\frac{1}{b^{2}}-\frac{g\left(r\right)}{r^{2}}+\frac{2g\left(r\right)\mathcal{L}}{L^{2}}\right)=:V_{eff},
    	\end{equation} 
    	where $b=L/E$ is impact parameter. Introduced new function $V_{eff}$ is called 'effective potential'.
    	
    	Now consider the motion of photon. For photon, $\mathcal{L}=0$, Eq.~\eqref{eq4_8} becomes
    	\begin{equation}\label{eq4_9}
    		\left(\frac{dr}{d\phi}\right)^{2}=r^4\left(\frac{1}{b^{2}}-\frac{g\left(r\right)}{r^{2}}\right).
    	\end{equation}
    	With the transform $u=1/r$, Eq.~\eqref{eq4_9} is rewritten as
    	\begin{equation}\label{eq4_10}
    		\left(\frac{du}{d\phi}\right)^{2}=G\left(u,b\right),
    	\end{equation}
    	where
    	\begin{equation}\label{eq4_11}
    		G\left(u,b\right)=\frac{1}{b^2}-u^2\left(1-2Mu+Q^2u^2-\frac{aQ^4}{20}u^6-u\alpha \ln\left(u\left|\alpha\right|\right)\right).
    	\end{equation}
    	For photon, parameter $b$ is the distance from the black hole to the asymptotic line of photon orbit at infinity.
    	
    	We first focus on the circular orbit of photon, which satisfies $V_{eff}=V_{eff}'=0$. Record the solution as $r=r_{ph}$, $b=b_{c}$. $r_{ph}$ is the radius of circular orbit and $b_{c}=r_{ph}/\sqrt{g\left(r_{ph}\right)}$ is impact parameter of the photon in circular orbit. Please notice that $A'$ represents $dA/dr$ for any scalar function $A$.
    	
\begin{figure}[htbp]
    		\centering
    		\includegraphics[width=0.6\textwidth]{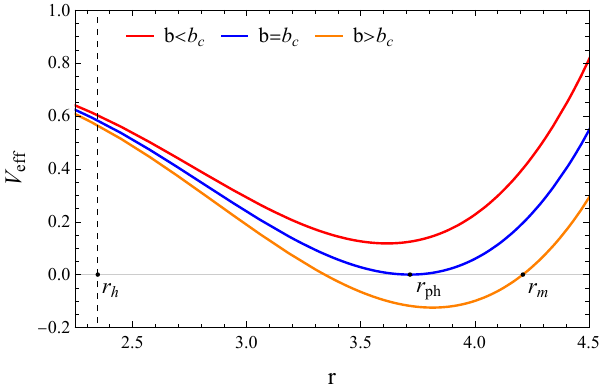}
    		\caption{Variation of $V_{eff}$ with respect to $r$. The red, blue and orange lines represents $b<b_{c}$, $b=b_{c}$ and $b>b_{c}$ respectively. The black dashed line represents event horizon. $r_h$ is radius of horizon, $r_{ph}$ is radius of circular orbit and $r_m$ is the maximum positive root of $V_{eff}$ for $b>b_{c}$. We set $Q=a=1, \alpha=-0.5$.}\label{Veff}
    	\end{figure}

    	Next, we will show how the value of $b$ influences the light's motion. Take $M=Q=a=1, \alpha=-0.5$ for an example, the function $V_{eff}\left(r\right)$ for different $b$ is plotted in Fig.~\ref{Veff}.
    	As seen in the figure:
    	
    	For $b>b_c$, there is a maximum positive root $r_m\left(b\right)$ of $V_{eff}$. The photon comes from infinity and will never enter the region $r<r_m$. Instead, it will reach the point $r_m$ and then return to infinity along a path with the same shape as the incoming trajectory;
    	For $b=b_c$, the photon will arrive $r=r_{ph}$ and then perpetually undergo circular motion;
    	For $b<b_c$, the photon will continuously approach the black hole until falling into the horizon. Consider an observer at infinity to ensure he will only receive parallel light, then $b_c$ is precisely the radius of the black hole shadow he observes, because of the reversibility of light path.
    	
        Define a function $\Phi_{b}\left(r\right)=\int_{0}^{\frac{1}{r}} \frac{1}{\sqrt{G\left(u,b\right)}}\,du$, then Eq.~\eqref{eq4_10} gives the total change of azimuth angle $\varphi$ for a certain orbit with the respect to $b$,
    	\begin{equation}\label{eq4_12}
			\varphi\left(b\right)=\begin{cases}
				\Phi_{b}\left(r_{h}\right), 
				& b<b_c,\\
				2\Phi_{b}\left(r_{m}\right), 
				& b>b_c.
			\end{cases}  		
    	\end{equation}
    	Obviously $\lim_{b \to \ b_{c}} \varphi\left(b\right)=+\infty$.
    	
\begin{figure}[htbp]
    	\centering
    	\includegraphics[width=0.9\textwidth]{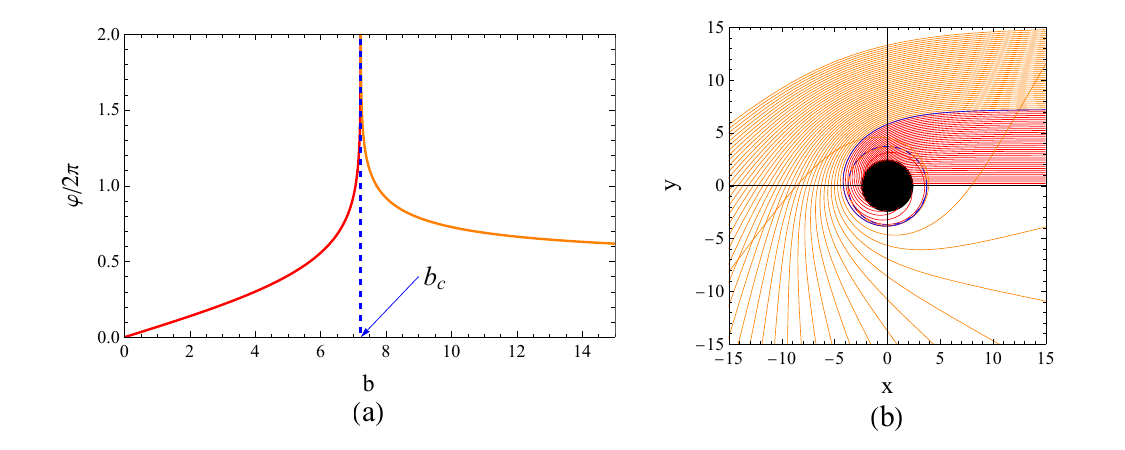}
    	\caption{(a) Graph of function $\varphi\left(b\right)$. (b) Photon orbits for different values of $b$. The red curves, blue curve and orange curves correspond to $b<b_c$, $b=b_c$ and $b>b_c$. The black hole is represented by the central black region in (b). We set $Q=a=1, \alpha=-0.5$.}\label{phi_b}
    \end{figure}

    	$\varphi\left(b\right)$ and its corresponding photon orbits are given in Fig.~\ref{phi_b}.
    	One clearly see that figure agrees with our analysis.
    	
    	Now go further to investigate how parameters $Q$, $a$ and $\alpha$ influence the motion of photon.
    \begin{figure}[htbp]
    	\centering
    	\includegraphics[width=1\textwidth]{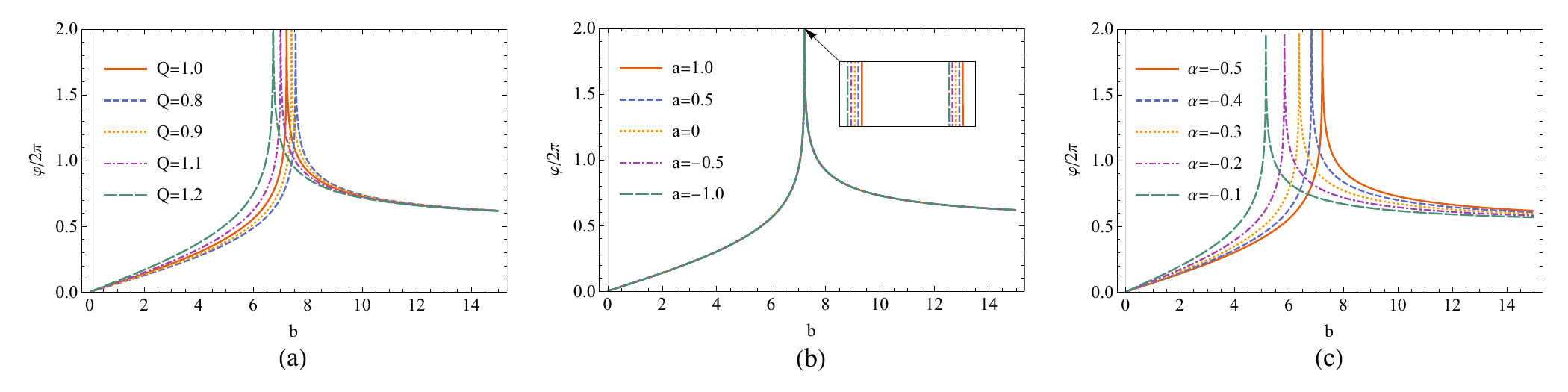}
    	\caption{Graphs of $\varphi\left(b\right)$ for different parameters $Q$, $a$ and $\alpha$. We set (a) $a=0$, $\alpha=-0.5$, (b) $Q=1$, $\alpha=-0.5$, (c) $Q=1$, $a=1$. Large version of (b) in region $b\in \left[7.2263,7.2313\right]$, $\varphi/2\pi\in\left[1.9995,2\right]$ is shown in the upper right corner.}\label{photon_orbit}
    \end{figure}
    	Fig.~\ref{photon_orbit} gives $\varphi\left(b\right)$ for different parameters. It shows the radius of shadow will decrease with the increase of magnitude of $\alpha$ and $Q$, and QED effect has almost no impact. It's necessary to point out the effects of $Q$ and $a$ agree with~\cite{QED_EH}, so we will only pay attention to the influence of PFDM in next subsection.
    \subsection{Image of thin accretion disk}\label{sec4_2}
    	Due to black hole's strong gravity, even light cannot escape, so black hole itself can't be observed by optical instrument. But generally, there is a accretion disk around the black hole. Because of the high-speed rotation of matter on the accretion disk, the effects of friction, collisions, and compression will result in intense thermal radiation, which can be observed by external observers.
    	
    	Thin accretion disk means that accretion disk can be simply abstracted into a plane with negligible thickness, and the influence of accretion disk on space-time can be ignored.
    	
\begin{figure}[htbp]
    		\centering
    		\includegraphics[width=1\textwidth]{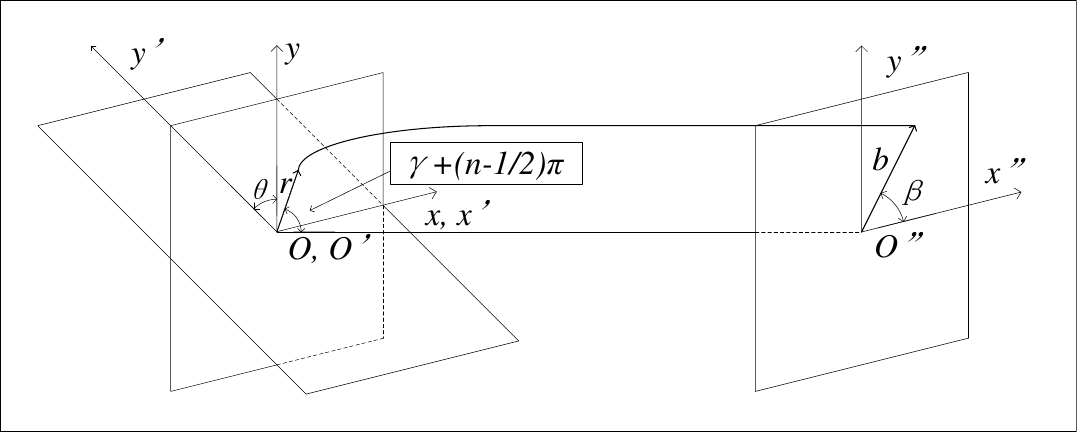}
    		\caption{Coordinate system.}\label{coordinate}
    	\end{figure}

    	Consider an observer at infinity to ensure he will receive parallel rays, which will simplify our discussion. Coordinate system is set in Fig.~\ref{coordinate}.
     
    	The center of black hole is located at point $O,O'$, accretion disk is in the $x'O'y'$ plane and observer is in $x''O''y''$ plane, which is used to show the image of disk. $xOy$ plane is parallel to $x''O''y''$ plane. The angle between $xOy$ plane and $x'O'y'$ plane is $\theta$. Right-hand Cartesian coordinate system is always built based on the plane to finish the calculation.
    	
    	Consider a photon located at point $\left(b,\beta\right)$ in $x''O''y''$ plane, departing in the vertical direction. The photon pass through the disk for $n$ times and finally arrive a point on the disk. The distance from this point to the center of black hole is $r$ and the change of photon's azimuth angle is $\gamma+\left(n-1/2\right)\pi$.
    	
        According to space analytic geometry,
        \begin{equation}\label{eq4_13}
    		\tan{\gamma}=\tan{\theta}\sin{\beta}.
        \end{equation}
        If fix $\theta$ and $n$, by calculating the photon's orbit one will obtain the value of $r$,
        \begin{equation}\label{eq4_14}
            r=r\left(b,\beta\right).
        \end{equation}
        If fix $r$, one will get a curve expressed by $b=b\left(\beta\right)$ in $xOy$ plane, which is precisely the image of circular orbit with a radius of $r$ on the disk. Because the photon in this case will pass through the disk for $n$ times, this image is called '$n^{\rm{th}}$ order image'.
    	
    	Now introduce a crucial concept: the inner edge $r_{in}$ of accretion disk. The inner edge $r_{in}$ of the accretion disk actually means the innermost stable circular orbit, because circular orbits within this range will be unstable, such that the particle will fall into the black hole or escape to flee farther when it is perturbed. $r_{in}$ should satisfy~\cite{shadow,isco}
    	\begin{equation}\label{eq4_15}
    		V_{eff}=V_{eff}'=V_{eff}''=0.
    	\end{equation}
    	
    	It's worth mentioning that according to~\cite{image1}, $n^{\rm{th}}$ order image is negligible when $n\geq3$. So we plot the first and second order images of accretion disk for different azimuth angles $\theta$ of observer.
    	  \begin{figure}[htbp]
    		\centering
    		\includegraphics[width=1\textwidth]{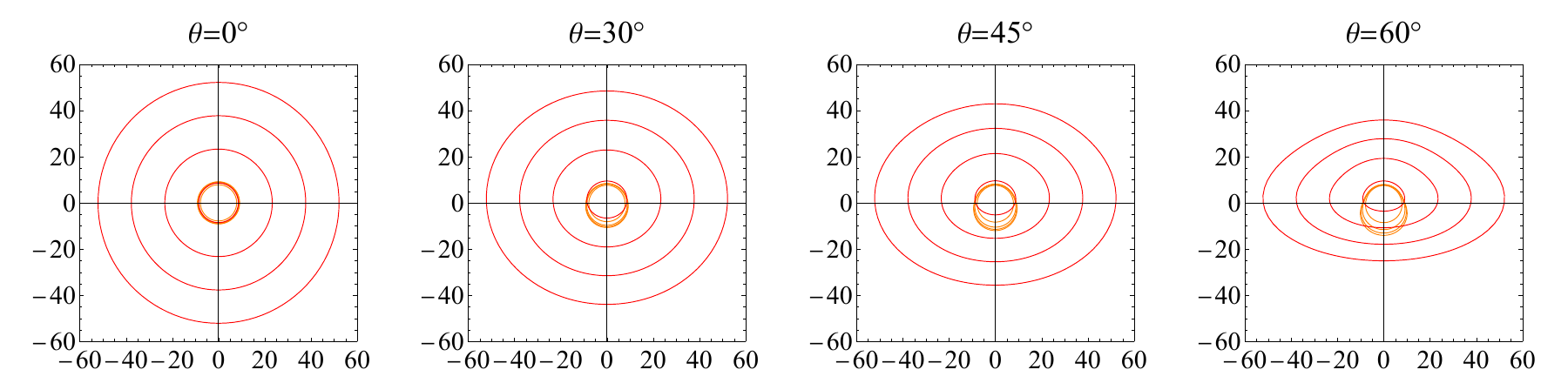}
    		\caption{The first order images (red lines) and second order images (orange lines) of circular orbit $r=\rm{constant.}$ on accretion disk. The values of $r$ are chosen from $r_{in}$ to $50M$ equidistantly. $\theta$ from left to right are $0$, $\pi/6$, $\pi/4$ and $\pi/3$ respectively. We set $Q=a=1$, $\alpha=-0.5$.}\label{accreation_disk_without_color}
    	\end{figure}
    	
    	As seen in Fig.~\ref{accreation_disk_without_color}, when the observer is in direction of North Pole, a series of circles will appear, which is consistent with the symmetry of space-time. As $\theta$ increases, the first order image gradually takes on a hat-like shape.
    	
    	To obtain the image with brightness, the Novikov-Thorne model is chosen~\cite{radiation_flux1,radiation_flux2},
    	\begin{equation}\label{eq4_16}
    		F\left(r\right)=-\frac{\mathcal{M}\Omega'}{4\pi\sqrt{-g}\left(E-\Omega L\right)^{2}}\int_{r_{in}}^{r} \left(E-\Omega L\right)L'dr,
    	\end{equation}
    	where $F(r)$ is radiation flux, $\mathcal{M}$ is black hole's accretion rate, $g$ is metric
    	determinant, and $r_{in}$ represents the inner edge of accretion
    	disk. $E$, $\Omega$ and $L$ are energy, angular velocity and angular momentum of physical particle moving on the disk.
    	
    	Consider a particle moving in a circular orbit, use $\mathcal{L}=-1/2$, one get $\Omega$ by particle's geodesic equation,
    	\begin{equation}\label{eq4_19}
    		\Omega=\frac{d\phi}{dt}=\pm\sqrt{-\frac{g_{tt}'}{g_{\phi\phi}'}}.
    	\end{equation}
    	Signs indicate two directions of rotation and have no essential distinctions. We choose positive in future calculations. Energy and angular momentum are given by Eq.~\eqref{eq4_3} and Eq.~\eqref{eq4_4}:
    	\begin{equation}\label{eq4_20}
    		E=\frac{-g_{tt}}{\sqrt{-g_{tt}-g_{\phi\phi}\Omega^{2}}},
    	\end{equation}
    	\begin{equation}\label{eq4_21}
    		L=\frac{g_{\phi\phi}\Omega}{\sqrt{-g_{tt}-g_{\phi\phi}\Omega^{2}}}.
    	\end{equation}

    	\begin{figure}[htbp]
    		\centering
    		\includegraphics[width=0.6\textwidth]{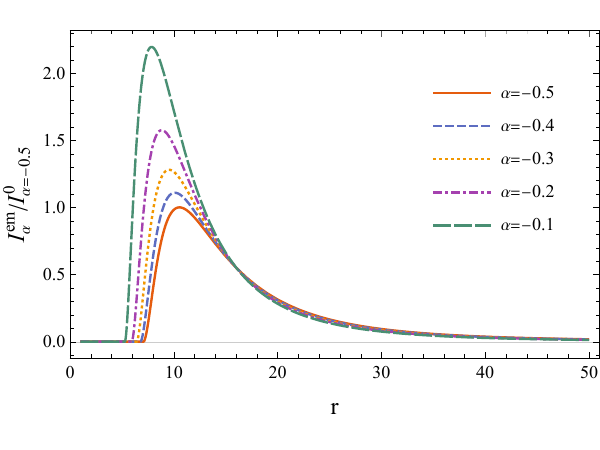}
    		\caption{Graphs of relative radiation flux for different $\alpha$. $I_{\alpha}^{em}$ is the radiation flux $F\left(r\right)$. $I_{\alpha=-0.5}^{0}$ is the maximum value of radiation flux for $\alpha=-0.5$. We set $Q=a=1$.}\label{I_em}
    	\end{figure}
    	Fig.~\ref{I_em} shows how PFDM affects the radiation flux of accretion disk. At short distance, the dark matter effect will significantly diminish the radiation flux. In contrast, at greater distance, the radiation flux will increase with the growth of $\alpha$.
    	
    	Due to the different gravitational fields at disk and observer and their relative motion, the frequency shift will happen. According to~\cite{image1}, the radiation flux observer received will be
    	\begin{equation}\label{eq4_22}
    		F_{obs}=\frac{F\left(r\right)}{\left(1+z\right)^{4}},
    	\end{equation}
    	where $z$ is redshift factor defined as
    	\begin{equation}\label{eq4_23}
    		1+z=\frac{E_{em}}{E_{obs}},
    	\end{equation}
    	$E_{em}$ and $E_{obs}$ are the the emitted light intensity at the light source and the light intensity received by the observer, respectively. According to~\cite{image2},
    	\begin{equation}\label{eq4_24}
    		E_{em}=-p_{\mu}u^{\mu},
    	\end{equation}
    	where $p_{\mu}=g_{\mu\nu}\dot{x^{\nu}}$ is photon's four-momentum and $u^{\mu}$ is particle's four-velocity. The received energy at infinity is naturally $E=-p_{t}$. According to geometry of Fig.~\ref{coordinate}, $p_{\phi}/p_{t}=b\sin \theta \cos \beta$, so
    	\begin{equation}\label{eq4_25}
    		1+z=u^{t}\left(1+\Omega\frac{p_{\phi}}{p_{t}}\right)=\frac{1+\Omega b\sin\theta\cos\beta}{\sqrt{-g_{tt}-g_{\phi\phi}\Omega^2}}.
    	\end{equation}
		Utilize above analysis one can get the image of the accretion disk with intensity distribution.
    	\begin{figure}[htbp]
    		\centering
    		\includegraphics[width=1\textwidth]{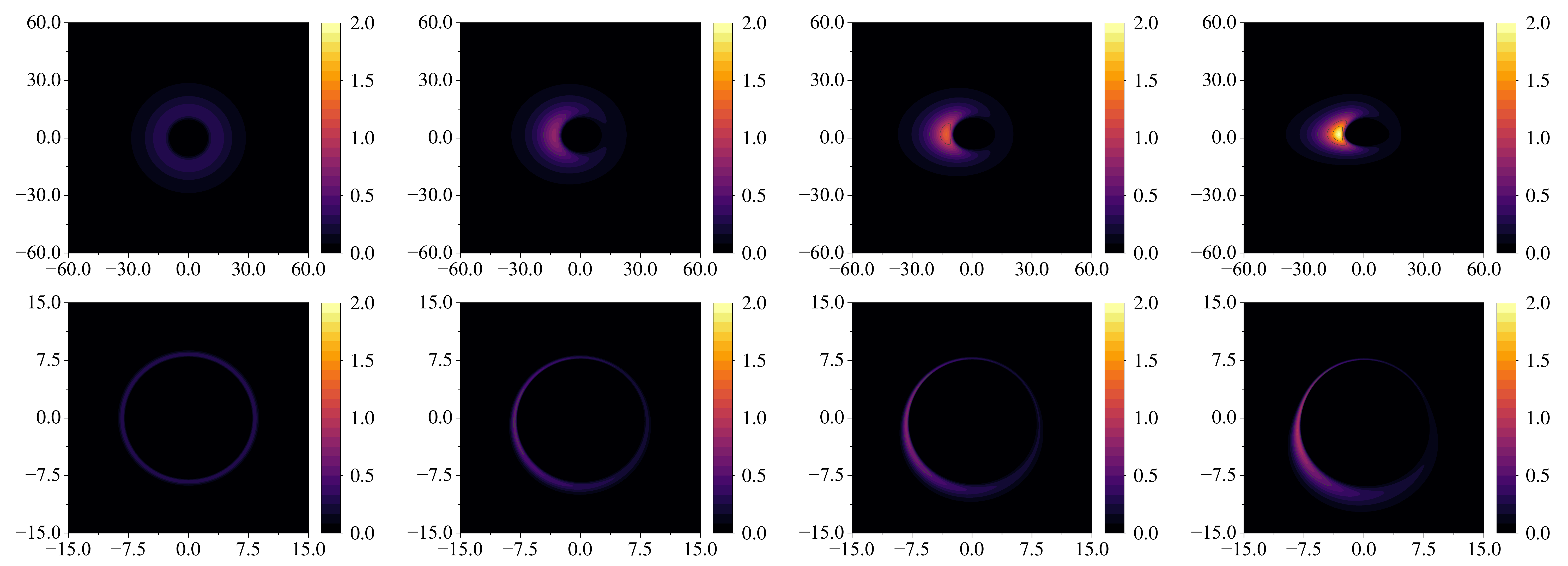}
    		\caption{First and second order images of accretion disk for different $\theta$. $\theta$ from left to right are $0$, $\pi/6$, $\pi/4$ and $\pi/3$ respectively. We set $Q=a=1$, $\alpha=-0.5$.}\label{I5}
    	\end{figure}
    	\begin{figure}[htbp]
    		\centering
    		\includegraphics[width=1\textwidth]{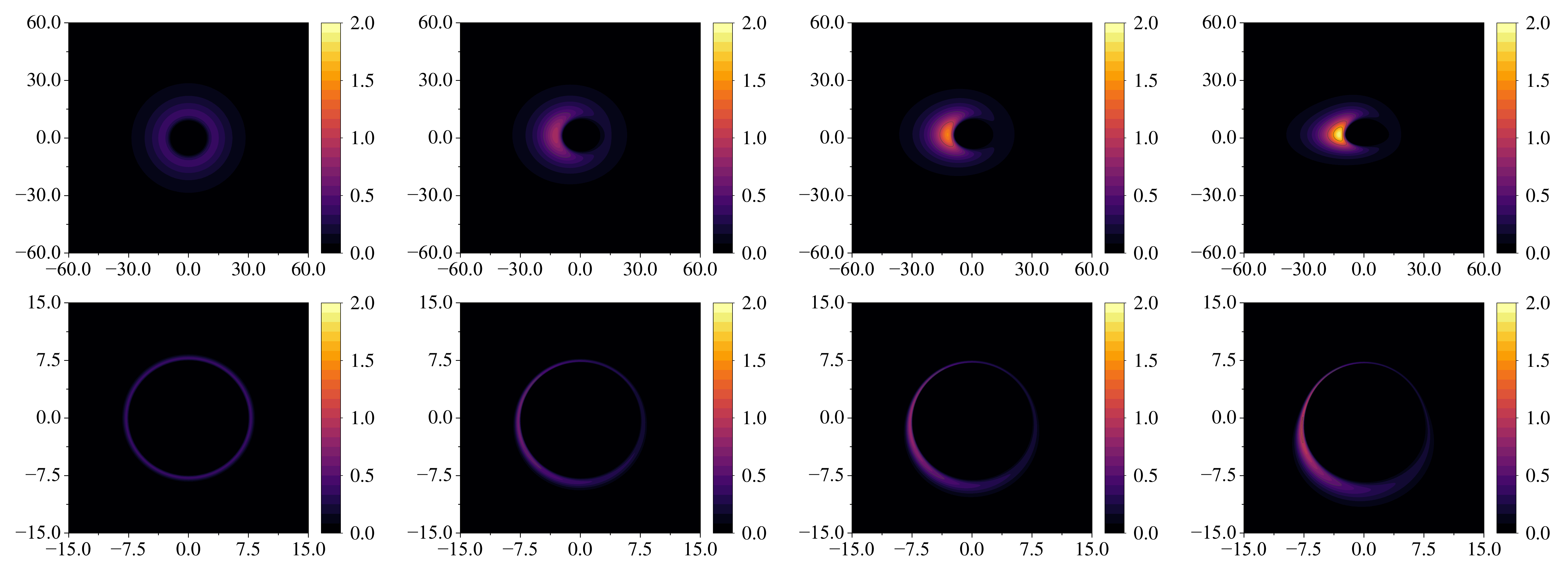}
    		\caption{First and second order images of accretion disk for different $\theta$. $\theta$ from left to right are $0$, $\pi/6$, $\pi/4$ and $\pi/3$ respectively. We set $Q=a=1$, $\alpha=-0.4$.}\label{I4}
    	\end{figure}
    	
    	The first and second order images with relative intensity distribution for different $\theta$ and dark matter parameters $\alpha$ is shown in Fig.~\ref{I5} and Fig.~\ref{I4}.
    	
    	As shown in images:
    	
    	(i) The shapes of image accord with Fig.~\ref{accreation_disk_without_color};
    	
    	(ii) Comparing to the first order image, the second order image already became very weak. It's foreseeable that for $n\geq3$, the image will be negligible;
    	
    	(iii) Doppler frequency shift is very evident. It cause the brightness on the left side is noticeably stronger than right side, because the matter on the left side of disk moves toward the observer;
    	
    	(iv) In general, the brightness of disk when $\alpha=-0.4$ is more strong than $\alpha=-0.5$. This conforms to the result in Fig.~\ref{I_em}, showing that the more intense dark matter effect will significantly reduce the brightness of disk. Although at a long distance, PFDM will increase the radiation flux slowly, the radiation flux rapidly tends to zero, making it appear negligible in the image. Of note is that the when $\theta=\pi/3$, the first image for $\alpha=-0.5$ is more intense than $\alpha=-0.4$, which is possible due to Doppler effect because redshift factor is also related to $\alpha$.
     
    \section{Thermodynamics}\label{sec5}
        In this section, we investigate the thermodynamics of EH black hole surrounded by PFDM in AdS space-time. Research for black hole thermodynamics in extended phase space requires us to regard cosmological constant $\Lambda$ as a thermodynamic variable. It seems to have contradiction, but there are also some good reasons why the variation of $\Lambda$ should be included in considerations~\cite{criticalexponent1}. And it's clear to see the physical meaning in~\cite{Kastor}, which shows that $\Lambda$ is regarded as pressure,
        \begin{equation}\label{eq5_1}
        	P=-\frac{\Lambda}{8\pi}.
        \end{equation}
    \subsection{EH-AdS black hole surrounded by PFDM}\label{sec5_1}
     	One can regard the cosmological constant $\Lambda$ as an effect of special Lagrangian written as
    	\begin{equation}\label{eq5_2}
    	\mathcal{L}^{\left(\Lambda\right)}=-\frac{\Lambda}{8\pi},
   		 \end{equation}
    	whose corresponding energy momentum tensor is 
    	\begin{equation}\label{eq5_3}
    	T_{\mu\nu}^{\left(\Lambda\right)}=-\frac{2}{\sqrt{-g}}\frac{\partial\sqrt{-g}\mathcal{L}^{\left(\Lambda\right)}}{\partial{g^{\mu\nu}}}=-\frac{\Lambda g_{\mu\nu}}{8\pi},
    	\end{equation}
    	which caused field equation with cosmological constant.
    	
        Anti-de Sitter space-time is
    	\begin{equation}\label{eq5_4}
        	ds^2_{AdS}=-\left(1-\frac{\Lambda r^2}{3}\right)dt^2+\frac{1}{\left(1-\frac{\Lambda r^2}{3}\right)}dr^2+r^2d\Omega^2.
        \end{equation}
        Its energy momentum tensor
        \begin{equation}\label{eq5_5}
        	T_{0}^{0\left(\Lambda\right)}=g^{0\alpha}T_{0\alpha}^{\left(\Lambda\right)}=-\frac{\Lambda}{8\pi},
        \end{equation}
        which apparently is just a constant without $f\left(r\right)$. And further, EH theory and model of PFDM have nothing to do with cosmological constant. That is to say, these three effects have no interactions.
        
        According to the discussion in Sect.~\ref{sec2}, when take the EH black hole surrounded by PFDM into AdS space-time, the line element will convert to
        \begin{equation}\label{eq5_6}
        	\begin{aligned}
        		ds^2&=-g\left(r\right)dt^2+\frac{1}{g\left(r\right)}dr^2+r^2d\Omega^2,\\
        		g\left(r\right)=1&-\frac{2M}{r}+\frac{Q^2}{r^2}-\frac{aQ^4}{20r^6}+\frac{\alpha}{r}\ln\left(\frac{r}{\left|\alpha\right|}\right)-\frac{\Lambda r^2}{3}.
        	\end{aligned}
        \end{equation}
	\subsection{Thermodynamic functions and equation of state}\label{sec5_2}
        The mass of black hole satisfies $g\left(r_{h}\right)=0$,
        \begin{equation}\label{eq5_7}
            M=\frac{r_{h}}{2}\left(1+\frac{\alpha}{r_{h}}\ln\left(\frac{r_{h}}{\left|\alpha\right|}\right)-\frac{\Lambda r_{h}^2}{3}+\frac{Q^2}{r_{h}^2}-\frac{aQ^4}{20r_{h}^6}\right).
        \end{equation}
        The Hawking temperature is defined by its surface gravity~\cite{TH},
        \begin{equation}\label{eq5_8}
            T_{H}=\frac{g'\left(r_{h}\right)}{4\pi}=\frac{1}{4\pi r_{h}}\left(1-\frac{Q^2}{r_{h}^2}+\frac{aQ^4}{4r_{h}^6}-\Lambda r_{h}^2+\frac{\alpha}{r_{h}}\right). 
        \end{equation}
         The entropy is
        \begin{equation}\label{eq5_9}
            S=\pi r_{h}^2.
        \end{equation}
        Thermodynamic first law is
        \begin{equation}\label{eq5_10}
            dM=TdS+VdP+\Phi dQ+\mu d\alpha.
        \end{equation}
        Here $M$ is mass of black hole, which is also interpreted as enthalpy. $T$ is temperature, $V$ is thermodynamic volume, $\Phi$ is electric potential and $\mu$ can be treated as dark matter potential.
        $V$, $\Phi$ and $\mu$ can be calculated form Eq.~\eqref{eq5_10},
        \begin{equation}\label{eq5_11}
            V=\left(\frac{\partial M}{\partial P}\right)_{S,Q,\alpha}=\frac{4\pi r_{h}^{3}}{3}.
        \end{equation}
        \begin{equation}\label{eq5_12}
            \Phi=\left(\frac{\partial M}{\partial Q}\right)_{S,P,\alpha}=\frac{Q}{r_{h}}-\frac{aQ^3}{10r_{h}^5}.
        \end{equation}
        \begin{equation}\label{eq5_13}
            \mu=\left(\frac{\partial M}{\partial \alpha}\right)_{S,P,Q}=\frac{1}{2}\left(\ln{\frac{r_{h}}{\left|\alpha\right|}}-1\right).
        \end{equation}
        It's crucial to verify whether $T=T_{H}$ because $T\neq T_{H}$ is possible~\cite{criticalexponent2}. If $T\neq T_{H}$, revising thermodynamic first law becomes necessary.
        
        For our black hole,
        \begin{equation}\label{eq5_14}
        	T=\left(\frac{\partial M}{\partial S}\right)_{P,Q,\alpha}.
        \end{equation}
        For arbitrary $A$,
        \begin{equation}\label{eq5_15}
        	\frac{\partial A}{\partial S}=\frac{1}{2\pi r_{h}}\frac{\partial A}{\partial r_{h}}.
        \end{equation}
        So
        \begin{equation}\label{eq5_16}
        	T=\frac{M'\left(r_{h}\right)}{2\pi r_{h}}=\frac{1}{4\pi r_{h}}\left(1-\frac{Q^2}{r_{h}^2}+\frac{aQ^4}{4r_{h}^6}-\Lambda r_{h}^2+\frac{\alpha}{r_{h}}\right)=T_{H}.
        \end{equation}
        Introduce the specific volume $v=2r_{h}$, equation of state can be solved by Eq.~\eqref{eq5_1}, Eq.~\eqref{eq5_8} and Eq.~\eqref{eq5_11},
        \begin{equation}\label{eq5_17}
            P=\frac{T}{v}-\frac{1}{2\pi v^{2}}+\frac{2Q^{2}}{\pi v^{4}}-\frac{8aQ^{4}}{\pi v^{8}}-\frac{\alpha}{\pi v^{3}}.
        \end{equation}
    \subsection{Phase transition and critical point}\label{sec5_3}  
        In the extended thermodynamic framework, the temperature $T$ is rewritten as a function of pressure $P$ and radius of horizon $r_{h}$, or entropy $S$
        \begin{equation}\label{eq5_18}
            T=\frac{1}{4\pi r_{h}}\left(1-\frac{Q^2}{r_{h}^2}+\frac{aQ^4}{4r_{h}^6}+8\pi P r_{h}^2+\frac{\alpha}{r_{h}}\right).
        \end{equation}

        \begin{figure}[htbp]
            \centering
            \includegraphics[width=0.95\textwidth]{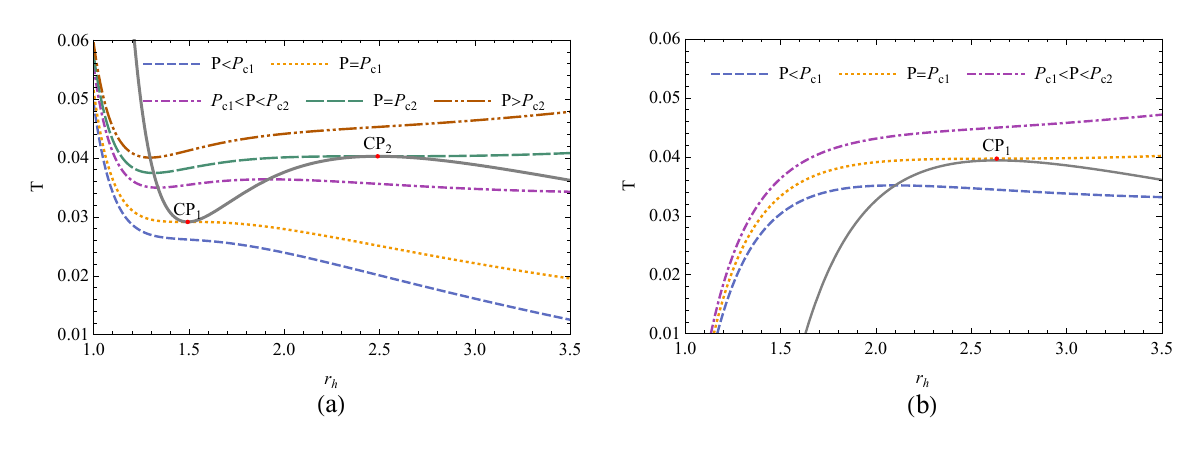}
            \caption{The solid lines in the figure represent second-order phase transition temperature $T_{p}$, the other lines represent isobaric temperature curve. $CP_ {1}$ and $CP_ {2}$ are the critical points. We set $Q=1,\alpha=-0.1$ and (a) $a=3$, (b) $a=-1$.}\label{T}
        \end{figure}
        We mainly pay attention to criticality, the condition of critical point is mentioned in~\cite{critical}, 
        \begin{equation}\label{eq5_24}
        	\left(\frac{\partial T}{\partial S}\right)_{P}=\left(\frac{\partial^2 T}{\partial S^2}\right)_{P}=0.
        \end{equation}
        And one can find the condition of second-order phase transition
        \begin{equation}\label{eq5_20}
            \left(\frac{\partial T}{\partial S}\right)_{P,\alpha,a,Q}=0,
        \end{equation}
        which is rewritten as following formula according to Eq.~\eqref{eq5_15},
        \begin{equation}\label{eq5_21}
            \left(\frac{\partial T}{\partial r_{h}}\right)_{P,\alpha,a,Q}=0,
        \end{equation}
        whose solution is
        \begin{equation}\label{eq5_22}
        	P=\frac{1}{32\pi r_{h}^{8}}\left(7aQ^{4}-12Q^{2}r_{h}^{4}+8\alpha r_{h}^{5}+4r_{h}^{6}\right).
        \end{equation}
        Substitute this solution into Eq.~\ref{eq5_18}, one could get the temperature when second-order phase transition occurs $T_{p}$ :
        \begin{equation}\label{eq5_23}
        	T_{p}=\frac{1}{4\pi r_{h}^{7}}\left(2aQ^{4}-4Q^{2}r_{h}^{4}+3\alpha r_{h}^{5}+2r_{h}^{6}\right).
        \end{equation}
        With the help of Appendix.~\ref{Appendix3}, critical point meets with
        \begin{equation}\label{eq5_25}
        	\frac{\partial T_{p}}{\partial r_{h}}=0,
        \end{equation}
        In our subsequent discussion, 'phase transition' we talk about default to 'second-order phase transition'.
        
        One can find the phase transition curves in Figure.~\ref{T}. One could find critical points $CP_1$ and $CP_2$ on the phase transition curves.
        
         \begin{figure}[htbp]
        	\centering
        	\includegraphics[width=0.95\textwidth]{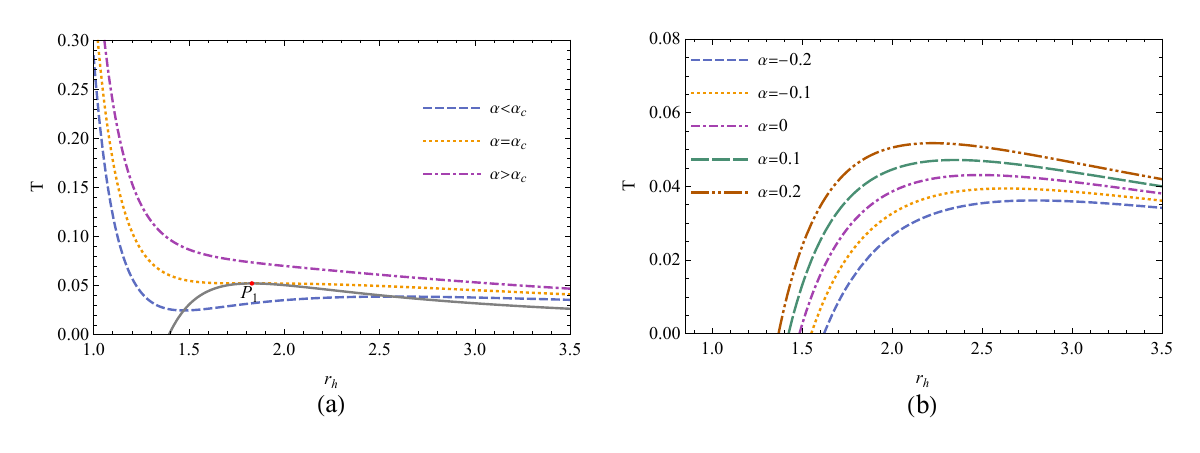}
        	\caption{The solid lines in figure represent critical point curve $T_{cp}$, the other lines represent phase transition curve $T_{p}$. $\alpha$ reaches its critical value at $P_1$. We set $Q=1$ and (a) $a=3$, (b) $a=-1$.}\label{Tp}
        \end{figure}
     	One could get the expression of $\alpha$ and the critical point temperature $T_{cp}$,
        \begin{equation}\label{eq5_27}
            \alpha=\frac{1}{3r_{h}^5}\left(-7aQ^4+6Q^2r_{h}^4-r_{h}^6\right),
        \end{equation}
        \begin{equation}\label{eq5_28}
            T_{cp}=\frac{1}{4\pi r_{h}^7}\left(-5aQ^4+2Q^2r_{h}^4+r_{h}^6\right).
        \end{equation}
        It's easy to find in Figure.~\ref{Tp}, there is a critical value of parameter $\alpha$ corresponding to the point $P_1$, which make two critical points combine into one. $P_1$ should meet with
        \begin{equation}\label{eq5_29}
            \left(\frac{\partial T_{cp}}{\partial r_{h}}\right)_{a,Q}=0.
        \end{equation}
        Use above equations, $a$ and the temperature $T_{co}$ of the combination point are derived,
        \begin{equation}\label{eq5_30}
            a=\frac{r_{h}^4}{35Q^4}\left(6Q^2+r_{h}^2\right),
        \end{equation}
        \begin{equation}\label{eq5_31}
            T_{co}=\frac{4Q^2+3r_{h}^2}{14\pi r_{h}^3}.
        \end{equation}
        \begin{figure}[htbp]
            \centering
            \includegraphics[width=0.95\textwidth]{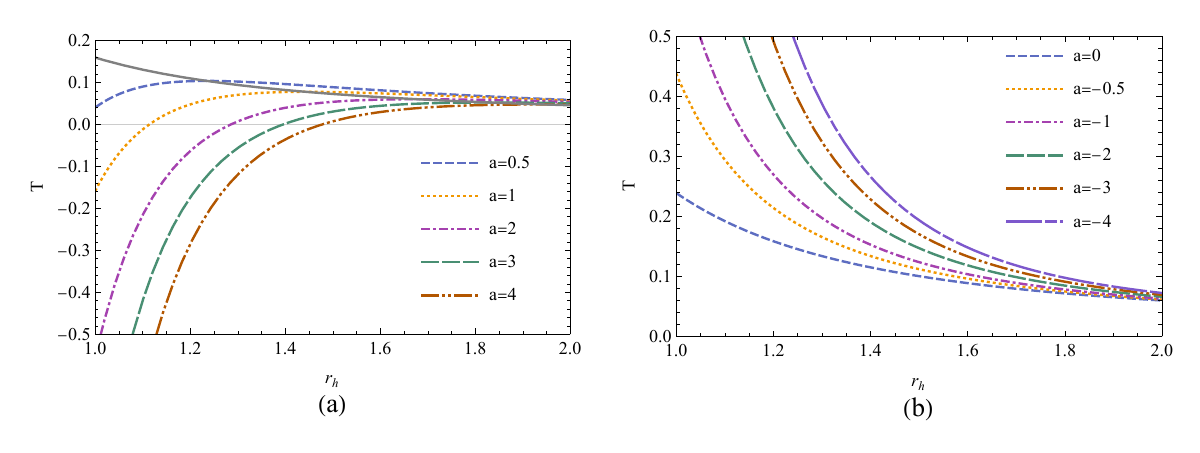}
            \caption{The solid lines in figure represent combination of critical points $T_{co}$, the other lines represent critical point curve $T_{cp}$. We set $Q=1$ and (a) $a>0$, (b) $a\leq0$.}\label{Tcp}
        \end{figure}
        
        As seen in Figure.~\ref{Tcp}, every critical point curve $T_{cp}$ has only one combination point for $a>0$, and there is only one positive real root of Eq.~\eqref{eq5_30},
        \begin{equation}\label{eq5_32}
            r_{h}=\sqrt{\frac{4Q^4}{A}+A-2Q^2},
        \end{equation}
        where
        \begin{equation}\label{eq5_33}
            A=\left(\frac{35aQ^4}{2}-8Q^6+\frac{\sqrt{35aQ^8\left(35a-32Q^2\right)}}{2}\right)^\frac{1}{3}.
        \end{equation}
      	One can get $\alpha_{c}$ by substituting Eq.~\eqref{eq5_32} into Eq.~\eqref{eq5_27}.
    \section{Conclusion and outlook}\label{sec6}
        Firstly, we derived a method to generate a new metric for static spherically symmetric space-time, which showed if multiple interactions have no coupling and energy momentum tensor $T_{0}^{0}$ is independent from metric to be determined, the solution of field equation will conform to the linear superposition. Using the this method, we obtained the metric of EH black hole surrounded by PFDM and study its optics and thermodynamics. It is found that shadow will increase when dark matter become intense. More strong dark matter will diminish the light intensity of accretion disk and there is exception when considering Doppler effect. In thermodynamics, one can see that PFDM will drastically affect the number of critical points when QED parameter is positive. 

        Looking forward, due to the conservation of angular momentum, black holes that are born in the collapse of stars are rarely non-rotating. Therefore, it will be more realistic to consider rotation. In thermodynamics, there are more properties to be researched, like critical exponents~\cite{criticalexponent1,criticalexponent2}. It is meaningful to study if thermodynamics of this kind of black holes is similar to Van der Waals system~\cite{RN-Ads1,RN-Ads2,Dolan}. And there is Joule-Thomson expansion, which is corresponding to isenthalpic process. It is meaningful to investigate Joule-Thomson expansion of this black hole compared to existed researches~\cite{JTE1,JTE2,JTE3}.
    \section*{Conflicts of interest}
        The authors declare that there are no conflicts of interest regarding the publication of this paper.
    \section*{Acknowledgments}
        We are grateful to Lei You, Yu-Cheng Tang and Yu-Hang Feng for their useful suggestions. We also thank the National Natural Science Foundation of China (Grant
        No.11571342) for supporting us on this work.
	\appendix
	\section{Equation of critical point}\label{Appendix3}
    \setcounter{equation}{0}
    \renewcommand{\theequation}{A.\arabic{equation}}
        The function $P\left(S\right)$ is defined to satisfy
        \begin{equation}\label{eqC1}
            \frac{\partial T}{\partial S}\left(S,P\left(S\right)\right)=0.
        \end{equation}
        Take the derivative of $S$ for both sides of Eq.~\eqref{eqC1},
        \begin{equation}\label{eqC2}
            \frac{\partial^{2} T}{\partial S^{2}}+\frac{\partial^{2} T}{\partial S \partial P}\frac{\partial P}{\partial S}=0.
        \end{equation}
        Considering $T_{p}\left(S\right)=T\left(S,P\left(S\right)\right)$,
        \begin{equation}\label{eqC3}
            \frac{\partial T_{p}}{\partial S}=\frac{\partial T}{\partial S}+\left(\frac{\partial T}{\partial P}\right)_{S}\frac{\partial P}{\partial S}=0.
        \end{equation}
        Use Eq.~\eqref{eqC2}, Eq.~\eqref{eqC3} and $\partial T/\partial S=0$, one get
        \begin{equation}\label{eqC4}
            \frac{\partial^{2} T}{\partial S^{2}}=-\frac{\partial^{2} T}{\partial P\partial S}\frac{1}{\left(\frac{\partial T}{\partial P}\right)_{S}}\frac{\partial T_{p}}{\partial S}.
        \end{equation}
        Clearly,
        \begin{equation}\label{eqC5}
            \frac{\partial T_{p}}{\partial S}=0\quad\Leftrightarrow\quad\frac{\partial^{2} T}{\partial S^{2}}=0,
        \end{equation}
        when $\partial T/\partial S=0$. ${\partial T_{p}}/{\partial S}=0$ also can be rewritten as ${\partial T_{p}}/{\partial r_{h}}=0$ by Eq.~\ref{eq5_15}.
		\bibliographystyle{unsrt}
		\bibliography{paper}
\end{document}